\documentclass[12pt,a4paper]{article}
\usepackage{graphicx}
\usepackage{amsmath}
\usepackage{amssymb}
\usepackage{epsfig}
\usepackage{times}

\begin{document}

\title{Einstein in 1916: ``On the Quantum Theory of Radiation"}

\author{Norbert Straumann, Uni Z\"{u}rich}

\maketitle

\subsection*{Introduction}
\label{intro}
A peak in Einstein's endeavor to extract as much information as possible about
the nature of radiation from the Planck distribution is his paper
``On the Quantum Theory of Radiation'' of 1916 \cite{Ein}.\footnote{The paper was first published in: Physikalische Gesellschaft Z\"{u}rich. Mitteilungen 18 (1916): 47-62. The same paper was received 3 March and published 15 March 1917 in: Physikalische Zeitschrift 18 (1917): 121-128.}
In the first part he gives a derivation
of Planck's formula which has become part of many textbooks on
quantum theory.

This part of the paper is now considered as the theoretical foundation of the laser,
 that was technically realized almost half a century later.

In the second part of his fundamental paper, Einstein discusses the
exchange of momentum between atoms and radiation by making
use of the theory of Brownian motion. Using a truly beautiful argument, 
which will be presented below, he shows that in every elementary process 
of radiation, and in particular in spontaneous emission, an amount $h\nu/c$
of momentum is emitted in a random direction and that the atomic system
suffers a corresponding recoil in the opposite direction.

This recoil was
first experimentally confirmed in 1933 by showing that a long and
narrow beam of excited sodium atoms widens up after spontaneous
emissions have taken place \cite{Fri}.
Einstein's paper \cite{Ein} ends with the following remarkable statement 
concerning the role of ``chance'' in his description of the
radiation processes by statistical laws, to which Pauli \cite{Pau}
drew special attention:
\begin{quote}
\textit{``The weakness of the theory lies, on the one hand, in the
fact that it does not bring us any closer to a merger with the
undulatory theory, and, on the other hand, in the fact that it
leaves the time and direction of elementary processes to `chance';
in spite of this I harbor full confidence in the trustworthiness
of the path entered upon.''}
\end{quote}

\subsection*{Derivation of the Planck distribution}
\label{sec:1}
In his novel derivation of the Planck distribution, Einstein
added the hitherto unknown process of
induced emission\footnote{Einstein's derivation shows that without
assuming a non-zero probability for induced emission one would
necessarily arrive at Wien's instead of Planck's radiation law.}, next 
to the familiar processes of spontaneous emission and induced
absorption. For each pair of energy levels he described the
statistical laws for these processes by three coefficients
(the famous $A$- and $B$-coefficients) and established two
relations between these coefficients on the basis of his earlier
correspondence argument in the classical Rayleigh-Jeans limit
and Wien's displacement law. In addition, the latter implies
that the energy difference $\varepsilon_n-\varepsilon_m$ between
two internal energy states of the atoms in equilibrium with thermal
radiation has to satisfy Bohr's frequency condition:
$\varepsilon_n-\varepsilon_m=h\nu_{nm}$. 

Einstein was very pleased by this derivation, about
which he wrote on August 11, 1916 to Besso: ``An amazingly simple
derivation of Planck's formula, I should like to say \emph{the}
derivation'' (CPAE, Vol.\,8, Doc.\,250).

In Dirac's 1927 radiation
theory these results follow -- without any correspondence
arguments -- from first principles.

\subsection*{Brownian  motion of an atom in the thermodynamic radiation field}
\label{sec:2}
In the second part, Einstein regarded in his paper as more important, he investigates the exchange of momentum between atoms and radiation by studying the  Brownian motion of molecules in the thermodynamic radiation field. A molecule experiences thereby two types of forces:

$\bullet$ A systematic drag force $Rv$, where $v$ is the velocity of the molecule, that leads in a small time interval $(t,t+\tau)$ to the momentum change $Rv\tau$.

$\bullet$ An irregular change of momentum $\Delta$ in the time $\tau$, due to fluctuations
of the radiation pressure.

In thermal equilibrium the following equation holds:
\[\langle(Mv-Rv\tau+\Delta)^2\rangle=\langle(Mv)^2\rangle. \]
Assuming that $\langle v\cdot\Delta\rangle=0$ we obtain 
the important relation, repeatedly used by Einstein in his earlier work on Brownian motion,
\begin{equation}
\langle\Delta^2\rangle=2R\langle M v^2\rangle\tau=2RkT\tau. 
\label{Eq:1}
\end{equation}
\paragraph{Calculation of the friction $R$.}
The computation of $R$ is not simple. For interested readers we present the details of Einstein's instructive calculation, thereby including some flesh of his profound investigation. (Others may prefer to jump directly to the simple result, given in equation (\ref{Eq:4}) below.)

The moving atom (molecule) sees an anisotropic radiation field. The friction force is the result of absorption and induced emission processes\footnote{The spontaneous emission has in the average no preferred direction, and therefore does in the average not transmit momentum to the atom.}. Let $K$ be the rest system of the radiation field, with respect to which the radiation intensity is isotropic. With $K'$ we denote the rest system of the atom. Let $S$ denote the partition sum
\[ S:=g_ne^{-E_n/kT}+g_me^{-E_m/kT}+\cdot\cdot\cdot~\]
for the energies $E_n$ and degeneracies $g_n$ of the atom. The fraction of time in the state $n$ is equal to $g_ne^{-E_n/kT}/S$, and similarly for the state $m$. Therefore, the number of absorptions $n\rightarrow m$ per unit time
into the solid angle $d\Omega'$ is equal to
\[\frac{1}{S}g_ne^{-E_n/kT}B^n_m\rho'_{\nu_0}\frac{d\Omega'}{4\pi},
~~~\nu_0=\frac{E_m-E_n}{h}\,,\] 
where $\rho_{\nu}$ denotes the energy density of the radiation as a function of frequency $\nu$.
Correspondingly, for induced emission $m\rightarrow n$ the number is
\[\frac{1}{S}g_me^{-E_m/kT}B^m_n\rho'_{\nu_0}\frac{d\Omega'}{4\pi}.\]
With Einstein's relation $g_mB^m_n=g_nB^n_m$ from the first part of the paper, the momentum transfer per
unit time in the $x$-direction is
\begin{equation*}
-Rv=\frac{h\nu_0}{c}\frac{1}{S}g_nB^n_m\left(e^{-E_n/kT}-e^{-E_m/kT}\right)
\times\int\rho'_{\nu_0}(\theta',\varphi')\cos\theta'\frac{d\Omega'}{4\pi}\,.
\label{Eq:2}
\end{equation*}
\paragraph{Calculation of $\rho_{\nu_0}(\theta',\varphi')$.}
With elementary calculations one can prove the following Lorentz invariance
\[\frac{\rho'_{\nu'}}{\nu'^3}=\frac{\rho_\nu}{\nu^3}\,.\]
(This follows more directly from the fact that $\rho(\nu)$ is proportional to $\nu^3$ times the Lorentz invariant distribution function $f$.) Using also the Doppler shift to first order in $v/c$: 
$\nu=\nu'\left(1+\frac{v}{c}\cos\theta'\right)$. This gives
\[
\rho'_{\nu'} \simeq \left(1+\frac{v}{c}\cos\theta'\right)^{-3}\rho_{[1+\frac{v}{c}\cos\theta']\nu'}~,\]
thus
\begin{eqnarray*}
\rho'_{\nu_0}&\simeq&\left[\rho_{\nu_0}+\left(\frac{\partial\rho_\nu}{\partial\nu}\right)_{\nu_0}
\nu_0\frac{v}{c}\cos\theta'\right]\left(1-3\frac{v}{c}\cos\theta'\right)\\
&\simeq&\rho_{\nu_0}+\frac{v}{c}\cos\theta'\left[\nu_0\left(\frac{\partial\rho_\nu}{\partial\nu}\right)_{\nu_0}
-3\rho_{\nu_0}\right]\,.
\end{eqnarray*}
To first order in $v/c$ we obtain
\begin{equation*}
\int\rho'_{\nu_0}\cos\theta'\frac{d\Omega'}{4\pi}=-\frac{v}{c}\left[\rho_{\nu_0}-\frac{1}{3}
\nu_0\left(\frac{\partial\rho_\nu}{\partial\nu}\right)_{\nu_0}\right]
=\frac{v}{c}\left\{\frac{\nu^4}{3}\frac{\partial}{\partial\nu}\left(\frac{\rho_\nu}{\nu^3}\right)\right\}_{\nu=\nu_0}.
\end{equation*}
Writing $\nu$ instead of $\nu_0$, we arrive at
\begin{equation*}
R=\frac{h\nu}{c^2}\frac{1}{S}g_nB^n_me^{-E_n/kT}\left(1-e^{-h\nu/kT}\right)
\times\left[-\frac{\nu^4}{3}\frac{\partial}{\partial\nu}\left(\frac{\rho_\nu}{\nu^3}\right)\right].
\end{equation*}
For the Planck distribution we obtain	
\begin{equation}
R=\frac{1}{3}\left(\frac{h\nu}{c}\right)^2\frac{1}{kT}\frac{g_n}{S}e^{-E_n/kT}B^n_m\rho_\nu\,.
\label{Eq:3}
\end{equation}
Note that $(g_n/S)\exp (-E_n/kT)B^n_m\rho_\nu$ is the number of
absorptions per unit time. Since this is equal to 1/2 times the number $Z$ of
elementary processes (induced and spontaneous emissions plus absorptions)
per unit time, we find for the friction force the simple result:
\begin{equation}
R=\frac{1}{3}\left(\frac{h\nu}{c}\right)^2\frac{1}{2kT}Z.
\label{Eq:4} 
\end{equation}
\paragraph{Interpretation}
Using this in (\ref{Eq:1}) we arrive at the crucial relation
\begin{equation}
\langle\Delta^2\rangle=\left(\frac{h\nu}{c}\right)^2\langle\cos^2\vartheta\rangle_{S^2} Z\tau \,.
\label{Eq:5} 
\end{equation}

From classical electrodynamics (and experience) we know that to a pencil of light with energy $\varepsilon$ belongs a momentum $\varepsilon/c$ in the direction of the light beam. From this Einstein concluded that the atom receives in an absorption process the momentum $(E_m-E_n)/c$ in the direction of the infalling radiation, and in an induced emission process the same momentum in the opposite direction. The crucial relation (\ref{Eq:5}) then implies that \emph{spontaneous} emission must also be directed in such a way that for every elementary process of radiation an amount $h\nu/c$ of momentum is emitted in a random direction, and that the atomic system suffers a corresponding recoil in the opposite direction. ``There is no radiation in spherical waves''\footnote{``Ausstrahlung in Kugelwellen gibt es nicht''.}, Einstein adds.

Einstein regarded this conclusion as the main result of his paper. In a letter to Michele Besso he states:
\begin{quote}
``\textit{With this, the existence of light-quanta is practically assured.}''\footnote{``Damit sind die Lichtquanten so gut wie gesichert'' (6. September 1916).}
\end{quote}
Two years later he added:
\begin{quote}
``\textit{I do not doubt anymore the \emph{reality} of radiation
quanta, although I still stand quite alone in this conviction.}''
\end{quote}

We conclude with the following interesting remark of Einstein to O. Stern:
\begin{quote}
``\textit{Ich habe hundertmal mehr \"{u}ber Quantenprobleme
nachgedacht, als \"{u}ber die allgemeine Relativit\"{a}tstheorie.}''
\end{quote}
%


\end{document}